\documentclass[a4paper,preprintnumbers,floatfix,superscriptaddress,pra,twocolumn,showpacs,notitlepage,longbibliography, aps]{revtex4-2}

\usepackage[utf8]{inputenc}
\usepackage[T1]{fontenc}
\usepackage{amsmath, amsthm, amssymb,amsfonts,mathbbol,amstext}
\usepackage{graphicx}
\usepackage{dcolumn}
\usepackage{bm}
\usepackage{bbm}
\usepackage{hyperref}
\usepackage{mathtools}
\usepackage{comment}
\usepackage{color}
\usepackage{multirow}
\usepackage{pgf,tikz}
\usepackage{mathrsfs}
\usepackage{physics}
\usetikzlibrary{arrows}
\usepackage{soul,xcolor}
\usepackage{adjustbox}
\usepackage{dsfont}

\begin{document}

\title{Quantum state texture of dynamical criticality}

\author{Lucas C. C\'eleri\href{https://orcid.org/0000-0001-5120-8176}{\includegraphics[scale=0.05]{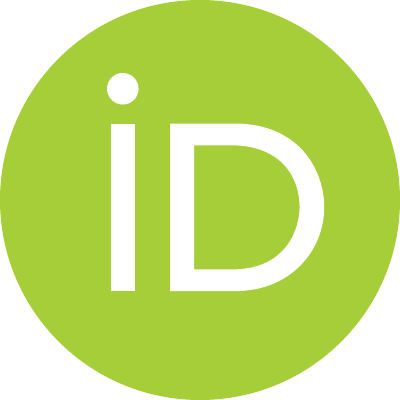}}}
\email{lucas@qpequi.com}
\affiliation{QPequi Group, Institute of Physics, Federal University of Goi\'as, Goi\^ania, Goi\'as, 74.690-900, Brazil}
\affiliation{São Carlos Institute of Physics, University of Sao Paulo, IFSC –USP,13566-590, São Carlos, SP, Brazil}
\author{Krissia Zawadzki\href{https://orcid.org/0000-0002-6133-0850}{\includegraphics[scale=0.05]{orcidid.pdf}}}
\affiliation{São Carlos Institute of Physics, University of Sao Paulo, IFSC –USP,13566-590, São Carlos, SP, Brazil}

\author{Ivan Medina\href{https://orcid.org/0000-0001-6428-2826}{\includegraphics[scale=0.05]{orcidid.pdf}}}
\affiliation{São Carlos Institute of Physics, University of Sao Paulo, IFSC –USP,13566-590, São Carlos, SP, Brazil}

\author{Diogo O. Soares-Pinto\href{https://orcid.org/0000-0002-4293-6144}{\includegraphics[scale=0.05]{orcidid.pdf}}}
\affiliation{São Carlos Institute of Physics, University of Sao Paulo, IFSC –USP,13566-590, São Carlos, SP, Brazil}

\begin{abstract}
We investigate the role of quantum state texture in dynamical quantum phase transitions by establishing a direct connection between critical nonequilibrium dynamics and the recently introduced notion of rugosity, a measure of quantum state texture. Considering a generic quench protocol, we analyze both standard formulations of dynamical quantum phase transitions. For type-I transitions, characterized through the long-time behavior of a dynamical order parameter, we show that the time-averaged rugosity, evaluated in the eigenbasis of the pre-quench Hamiltonian, exhibits an order-parameter-like behavior across the transition. In the Lipkin-Meshkov-Glick model, this behavior is traced back to the underlying semiclassical structure, where crossing the excited-state quantum phase transition separatrix controls the redistribution of the quantum state over the pre-quench energy basis. We further demonstrate that the response function associated with the time-averaged rugosity develops a pronounced critical peak whose maximum grows algebraically with the system size, providing quantitative finite-size evidence of its sensitivity to the dynamical critical point. For type-II transitions, characterized by nonanalyticities in the Loschmidt rate function, we establish an exact connection with rugosity. For a suitable choice of basis, the rate function is exactly given by the density of rugosity, establishing a model-independent equivalence. Moreover, we show that even in physically motivated bases, such as the pre-quench energy eigenbasis, rugosity provides clear signatures of dynamical criticality. Our results establish rugosity as a sensitive probe of dynamical quantum phase transitions while highlighting its role as a basis-dependent measure of quantum state texture, thereby providing an information-theoretic perspective on nonequilibrium quantum critical phenomena.
\end{abstract}

\maketitle

\section{Introduction}

Studying the properties of quantum resources is a fundamental task not only for deepening our understanding of the structure of quantum theory, but also for enabling technological applications that exploit genuinely quantum features. In this direction, resource theories have emerged as a unifying and systematic framework to identify, quantify, and manipulate such resources under physically motivated constraints~\cite{Chitambar2019}. Among the various proposals, some resource theories have reached a particularly mature stage, including those of entanglement~\cite{Horodecki2009}, stabilizerness~\cite{Veitch2014}, quantum coherence~\cite{Streltsov2017} and quantum thermodynamics~\cite{Lostaglio2019}. These frameworks have led to significant progress on both the theoretical side, by clarifying the operational meaning of nonclassical features, and the experimental side, by guiding the design and certification of quantum protocols. Their impact spans a wide range of fields, from quantum computation and communication~\cite{Hahn2025,Son2025} to many-body physics~\cite{Garcia2025} and thermodynamics~\cite{Brandao2013}, thus explicitly demonstrating the central role of resource-theoretic thinking in modern quantum science.

More recently, a new resource theory has been introduced to characterize the structural irregularities of quantum states, establishing a direct connection between the geometry of density matrices and an operationally meaningful resource, the so-called rugosity~\cite{Parisio2024}. This quantity is intrinsically basis dependent, in close analogy with quantum coherence, but it captures a finer notion of structure by being sensitive not only to the magnitude of off-diagonal elements but also to their relative phases. In this sense, rugosity provides a measure of quantum state texture, quantifying how far a given state is from a completely uniform (flat) superposition in a chosen basis. Beyond its original formulation, several recent works have proposed alternative measures of quantum texture, explored their geometric and information-theoretic interpretations, and connected them to broader notions of complexity and structure in quantum states~\cite{Cao2026,Liu2025,Cui2026,Chen2026,Muthuganesan2026}. Applications have already begun to emerge in different contexts, including equilibrium quantum phase transitions~\cite{Patra2026} and relativistic open quantum systems~\cite{Huang2025}, suggesting that texture-based quantities may provide a new lens through which to analyze quantum phenomena.

In this work, we investigate the texture of the quantum state in the context of dynamical quantum phase transitions (DQPTs)~\cite{Heyl2013,Heyl2018,Heyl2019}. Unlike equilibrium phase transitions, which are governed by thermal ensembles and free-energy landscapes, DQPTs arise in the unitary time evolution of isolated quantum many-body systems following a quench. In such situations, the system typically explores a broad set of excited eigenstates with a highly nonthermal distribution, and its dynamics cannot be described within the standard framework of equilibrium statistical mechanics. This makes nonequilibrium dynamics a particularly rich setting for a probe of fundamental questions about ergodicity, thermalization, and the emergence of macroscopic behavior from microscopic quantum dynamics. In particular, the study of DQPTs has revealed that sharp, nonanalytic behavior, reminiscent of equilibrium critical phenomena, can emerge purely in time, providing a dynamical notion of criticality.

Recent developments have highlighted deep connections between DQPTs and concepts from quantum information and quantum thermodynamics. For example, within a phase space description, it has been shown that the Wehrl entropy exhibits quasi-monotonic growth and reaches maximal production rates near the dynamical critical point~\cite{Goes2020}, signaling enhanced mixing and spreading in phase space. Similarly, Krylov complexity has been identified as an effective order parameter for dynamical criticality in certain models~\cite{Bento2024}, reflecting the rapid growth of operator complexity and the exploration of Hilbert space. In addition, dynamical criticality has been associated with an acceleration of thermodynamic entropy production~\cite{Nascimento2024}, as well as with the buildup of symmetry-related coherences, as captured by asymmetry measures~\cite{Nascimento2025}. Taken together, these results point toward a unified picture in which DQPTs act as highly efficient mechanisms for redistributing quantum information, linking entropy production, complexity growth, and symmetry properties in nonequilibrium dynamics.

Within this broader context, our goal is to investigate how quantum-state texture behaves under nonequilibrium evolution and whether rugosity can provide useful signatures of dynamical quantum phase transitions. For type-II DQPTs~\cite{Heyl2014,Heyl2015,Vajna2015,Heyl2017,Flaschner2018}, we show that, for any pure initial state, one may construct a basis in which that state is the corresponding flat state. In this representation, the rugosity of the time-evolved state is exactly minus the logarithm of the Loschmidt echo, so that the rugosity density coincides with the Loschmidt rate function, whose behavior characterizes dynamical criticality. This establishes the existence of a basis in which the two quantities coincide.

For type-I DQPTs~\cite{Sciolla2013,Zhang2017,Smacchia2015,Halimeh2017,Chen2020}, we focus on the Lipkin-Meshkov-Glick (LMG) model and evaluate rugosity in the eigenbasis of the pre-quench Hamiltonian, which is fixed independently of the subsequent dynamics. In this physically motivated basis, the long-time averaged rugosity exhibits an order-parameter-like change across the transition, while its associated response function develops a peak that becomes progressively higher and narrower with increasing system size. This behavior can be understood from the semiclassical structure of the LMG model, where crossing the excited-state quantum phase transition separatrix changes the redistribution and coherent organization of amplitudes over the pre-quench energy basis.

This article is organized as follows. In Sec.~\ref{sec:dqpt-rugosity}, we present a brief review of both types of dynamical quantum phase transition, together with the concept of quantum state texture and the definition of rugosity. In Sec.~\ref{sec:lmg}, we introduce the Lipkin-Meshkov-Glick model, which serves as the basis for our analysis. The main results are presented in Sec.~\ref{sec:results}, followed by a detailed discussion in Sec.~\ref{sec:discussion}. Finally, Sec.~\ref{sec:conclusion} is devoted to our concluding remarks.

\section{\label{sec:dqpt-rugosity} DQPT and rugosity}

Consider a quantum system described by a Hamiltonian $H(h)$, where $h$ denotes an externally controlled parameter or, more generally, a set of control parameters. A dynamical quantum phase transition is typically investigated through a sudden-quench protocol. The system is initially prepared in the ground state $\ket{\psi_0}$ of the Hamiltonian $H_0\equiv H(h_0)$. At time $t=0$, the control parameter abruptly changes from $h_0$ to $h_f$, and the subsequent evolution is governed by the post-quench Hamiltonian $H_f\equiv H(h_f)$. The state at time $t$ is therefore
\begin{equation}
\ket{\psi_t} = e^{-iH_f t}\ket{\psi_0}.
\end{equation}

Two complementary notions of dynamical quantum phase transitions have been developed in the literature. The first, commonly referred to as type-I (DQPT-I), is formulated in terms of the long-term behavior of an order parameter~\cite{Sciolla2013,Zhang2017,Smacchia2015,Halimeh2017,Chen2020}. Given an observable $O$, one defines its time average as
\begin{equation}
\overline{\langle O \rangle} = \lim_{T\to\infty} \frac{1}{T}\int_0^T \dd t\, \langle O\rangle,
\label{eq:order}
\end{equation}
where $\langle O\rangle=\bra{\psi_t}O\ket{\psi_t}$. A DQPT-I occurs when $\overline{\langle O\rangle}$ changes qualitatively as a function of the quench parameter. In many systems, including the Lipkin-Meshkov-Glick model considered here, the long-time order parameter is finite on one side of the transition and vanishes on the other. This behavior distinguishes a regime in which the dynamics retain memory of the initially symmetry-broken state from a regime in which the symmetry is dynamically restored.

The second characterization, known as type-II (DQPT-II), is based on the probability that the system returns to its initial state~\cite{Heyl2014,Heyl2015,Vajna2015,Heyl2017,Flaschner2018}. The latter is associated with the Loschmidt amplitude
\begin{equation}
\mathcal{G}_t = \bra{\psi_0}e^{-iH_f t}\ket{\psi_0},
\end{equation}
whose squared modulus
\begin{equation}
\mathcal{L}_t=|\mathcal{G}_t|^2
\end{equation}
is the Loschmidt echo, or return probability. In analogy with the free-energy density of equilibrium statistical mechanics, one defines the rate function
\begin{equation}
\lambda_t = -\frac{1}{N}\ln \mathcal{L}_t,
\label{eq:rate}
\end{equation}
where $N$ denotes the number of subsystems (for example, the number of spins in a spin chain). In the thermodynamic limit, a DQPT-II is signaled by nonanalytic behavior of $\lambda_t$, which typically appears as cusplike singularities at a sequence of critical times~\cite{Heyl2018,Heyl2019}. These singularities arise when Fisher zeros of the analytically continued Loschmidt amplitude cross the real-time axis, in close analogy with the role played by Lee-Yang and Fisher zeros in equilibrium phase transitions~\cite{Heyl2014,Heyl2015,Vajna2015,Heyl2017,Flaschner2018}.

The aim of this work is to relate this nonequilibrium critical behavior to a recently introduced quantum resource theory for the quantum state texture, which we describe now~\cite{Parisio2024}. Consider a finite-dimensional Hilbert space $\mathcal{H}_d$ of dimension $d$, along with an orthonormal basis $\{\ket{\epsilon_m}\}_{m=1}^{d}$. Associated with this basis one defines the state of minimum texture, or flat state
\begin{equation}
\ket{\omega} = \frac{1}{\sqrt d}\sum_{m=1}^d \ket{\epsilon_m}.
\end{equation}
This is a unique pure state whose amplitudes are completely uniform in the chosen basis. The texture of a general quantum state $\rho$ can be quantified by rugosity, which is defined as~\cite{Parisio2024}
\begin{equation}
R_{\epsilon}(\rho) = -\ln\left[\bra{\omega}\rho\ket{\omega}\right],
\label{eq:rugosity}
\end{equation}
with the subindex $\epsilon$ denoting the basis.

Rugosity quantifies how distinguishable $\rho$ is from the flat reference state. By construction, $R_{\epsilon}(\rho)\geq 0$, with equality if and only if $\rho=\ket{\omega}\bra{\omega}$. In contrast, $R_{\epsilon}(\rho)\rightarrow\infty$ whenever $\rho$ becomes orthogonal to $\ket{\omega}$. Therefore, large values of rugosity indicate that the state develops a highly nonuniform structure on the chosen basis~\cite{Parisio2024}.

The basis dependence of rugosity is essential for its physical interpretation. In the context of a DQPT (quench protocol), a natural choice is the eigenbasis of the pre-quench Hamiltonian,
\begin{equation}
H_0\ket{\epsilon_m}=\epsilon_m\ket{\epsilon_m},
\end{equation}
for which the flat state corresponds to an equal superposition of all initial energy eigenstates. Rugosity then measures the departure of the time-evolved state from the flat coherent superposition over the pre-quench spectrum. Importantly, this depends not only on the redistribution of populations among the eigenstates of $H_0$, but also on the relative phases of their amplitudes.

However, at first sight, this choice appears unrelated to the return probability defining the DQPT-II. The initial state is the ground state of the pre-quench Hamiltonian and is therefore highly localized in the corresponding energy eigenbasis rather than flat. Consequently, the Loschmidt echo is not, in general, directly related to the overlap with the flat state that defines rugosity. Nevertheless, for any pure initial state, one may always construct a basis in which the initial states itself becomes the corresponding flat state. In this representation, the rugosity of the time-evolved state reduces exactly to minus the logarithm of the Loschmidt echo and, therefore, its density coincides with the DQPT-II rate function. This establishes an exact and model-independent identity between rugosity and the Loschmidt rate function.

The situation is different for type-I dynamical quantum phase transitions. In this case, we do not engineer the basis to obtain a particular identity. Instead, we consider the physically motivated eigenbasis of the pre-quench Hamiltonian, which is fixed by the quench protocol. We show that, in the LMG model, the time-averaged rugosity computed in this basis exhibits an order-parameter-like behavior across the dynamical transition and that its associated response function develops a pronounced finite-size critical scaling. These two results highlight complementary roles of rugosity: an exact basis-dependent representation of DQPT-II and a physically motivated probe of type-I dynamical criticality in the LMG model.

In the following section, we introduce the specific model employed in this work in order to make this connection explicit.

\section{\label{sec:lmg}The model}

Let us consider the Lipkin-Meshkov-Glick Hamiltonian~\cite{Lipkin1965,Meshkov1965,Glick1965,Ribeiro2008,Ribeiro2007,Vidal2004} restricted to the fully symmetric sector,
\begin{equation}
H_{\mathrm{LMG}}=-hJ_x-\frac{\delta}{2j}J_z^2,
\label{eq:quantum_hamiltonian}
\end{equation}
where $h,\delta\geq0$, $J_\alpha=(1/2)\sum_{i=1}^{N}\sigma_\alpha^{(i)}$ are collective spin operators, and $N=2j$. Restricting the dynamics to the symmetric subspace fixes the total spin to its maximum value $j=N/2$, reducing the many-body problem to the dynamics of a single collective spin of length $j$. Despite its simplicity, the model exhibits equilibrium~\cite{Lipkin1965,Meshkov1965,Glick1965}, excited-state~\cite{Corps2022}, and dynamical~\cite{Goes2020} quantum phase transitions.

In the ferromagnetic regime ($\delta>0$) the transverse field term tends to align the collective spin along the $x$-direction, whereas the interaction term favors a finite magnetization along $z$. In the thermodynamic limit ($j\to\infty$) this competition gives rise to an equilibrium quantum phase transition at the critical field
\begin{equation}
h_c^{\mathrm{QPT}}=\delta.
\end{equation}
For $h>\delta$, the transverse field dominates and the ground state belongs to the symmetric phase, characterized by $\langle J_z\rangle=0$. By contrast, for $h<\delta$, the interaction term prevails, and the system enters a symmetry-broken phase with two degenerate minima carrying opposite magnetization,
\begin{equation}
\frac{\langle J_z\rangle}{j}=\pm\sqrt{1-\frac{h^2}{\delta^2}}.
\end{equation}
The equilibrium transition therefore corresponds to the bifurcation of the ground-state minimum into two symmetry-related minima.

With regard to the excited-state quantum phase transition (ESQPT), its physical origin becomes particularly clear in the semiclassical limit. Introducing the canonical variables $(z,\phi)$, defined by $z=J_z/j$ and $J_x=j\sqrt{1-z^2}\cos\phi$, the Hamiltonian density takes the form of
\begin{equation}
\mathcal{H}(z,\phi)=\frac{H_{\mathrm{LMG}}}{j}
=-h\sqrt{1-z^2}\cos\phi-\frac{\delta}{2}z^2.
\label{eq:classical_hamiltonian}
\end{equation}
For $h<\delta$, the classical energy landscape has the structure of a double well. The two minima are located at
\begin{equation}
z=\pm\sqrt{1-\frac{h^2}{\delta^2}}, \qquad \phi=0,
\end{equation}
whereas the point $(z=0,\phi=0)$ is no longer stable. Instead, it becomes a saddle point that separates the two wells. The energy of this saddle point plays a central role. Evaluating Eq.~\eqref{eq:classical_hamiltonian} at $(z,\phi)=(0,0)$, one obtains the critical energy
\begin{equation}
E_c^{\mathrm{ESQPT}}/j=-h.
\end{equation}
Classically, this energy defines the separatrix that divides two qualitatively distinct families of trajectories. For energies below $E_c^{\mathrm{ESQPT}}$, the motion remains confined to one of the two wells, so the sign of $J_z$ is preserved during evolution. For energies above the separatrix, the trajectories can cross between the two wells, dynamically restoring the symmetry. In the quantum problem, this change in the classical phase-space structure manifests itself as an excited-state quantum phase transition, signaled by singular behavior in the density of states and by a marked change in the structure of the excited eigenstates.

Let us now consider a sudden quench of the transverse field, $h_0\to h_f$, with the system initially prepared in the ground state of the ordered phase, $h_0<\delta$. After quenching, the state evolves under the final Hamiltonian $H_f$, but its energy with respect to this Hamiltonian is fixed by the initial condition. In the semiclassical limit, the injected energy is
\begin{equation}
E_{\mathrm{inj}}
=-\frac{\delta}{2}-h_f\frac{h_0}{\delta}.
\end{equation}
A dynamical quantum phase transition occurs when the injected energy coincides with the ESQPT separatrix energy of the post-quench Hamiltonian,
\begin{equation}
E_{\mathrm{inj}}=E_c^{\mathrm{ESQPT}}(h_f)=-h_f.
\end{equation}
Solving this condition yields the dynamical critical field
\begin{equation}
h_c^{\mathrm{DQPT}}=\frac{\delta+h_0}{2}.
\end{equation}
The dynamical phase transition therefore admits a simple semiclassical interpretation: it occurs precisely when the quench injects enough energy for the system to reach the saddle point separating the two wells. Below this threshold, the evolution remains confined to a single symmetry-broken sector, and the long-time magnetization stays finite. Above it, the dynamics explore both wells, leading to dynamical symmetry restoration and a vanishing long-time magnetization.

These considerations are directly relevant for the interpretation of rugosity. As discussed above, a natural choice of basis in the context of a quench is the eigenbasis of the pre-quench Hamiltonian. In the LMG model, this basis is particularly meaningful because the DQPT is controlled by competition between the two semiclassical sectors associated with the double-well structure. The pre-quench basis resolves these sectors, while the ESQPT separatrix determines whether the post-quench dynamics remain localized in one well or explore both.

Consequently, the rugosity evaluated in the pre-quench basis is expected to be highly sensitive to the crossing of the ESQPT separatrix and, therefore, to the DQPT itself. Nevertheless, the state that initiates the dynamics is not the flat state $\ket{\omega}$, but rather the ground state $\ket{\psi_0}$ of $H_0$. In general, $\ket{\psi_0}$ is strongly localized in the pre-quench energy basis and already possesses a finite rugosity at $t=0$. The subsequent evolution under $H_f$ redistributes both the populations and the phases on this basis, thereby modifying the rugosity over time.

The central question addressed in this work is whether the evolution of rugosity can encode signatures of dynamical quantum criticality. In the following sections, we address this question from the two complementary perspectives of dynamical quantum phase transitions. Within the framework of type-II DQPTs, we establish an exact relation between the Loschmidt rate function and the density of rugosity by constructing the basis in which the initial state is flat. Within the framework of type-I DQPTs, we investigate the physically motivated pre-quench energy eigenbasis and show that, in the LMG model, the time-averaged rugosity exhibits an order-parameter-like behavior, while its associated response function displays clear finite-size signatures of the dynamical transition. Together, these results demonstrate that rugosity provides a useful information-theoretic perspective on dynamical quantum criticality, while highlighting the distinct roles played by basis construction and basis selection.

\section{\label{sec:results}Results}

We first consider the type-I dynamical quantum phase transition, which is identified through the long-time behavior of the order parameter. 

Before discussing the numerical results, we briefly summarize the procedure used to generate the figures. For each value of the total angular momentum $j$, the dynamics is restricted to the fully symmetric sector, whose Hilbert-space dimension is $d=2j+1$. The system is initially prepared in one of the symmetry-broken ground states of the pre-quench Hamiltonian with $h_0=0$. At $t=0$, the transverse field is suddenly changed to a final value $h_f$, and the post-quench Hamiltonian $H_f$ is numerically diagonalized. The time-evolved state is them computed in this basis. At each time, we compute the longitudinal magnetization and the rugosity with respect to the pre-quench energy eigenbasis. The corresponding long-time averages are then obtained over an interval $T\simeq10^{4}\nu^{-1}$, where $\nu$ is the smallest Bohr frequency in the range of $h_f$ and $j$ considered. The calculation is repeated for all values of $h_f$ and $j$ shown in the figures.

In equilibrium statistical mechanics, continuous phase transitions are commonly characterized through the behavior of an order parameter and its associated response functions, whose singular behavior signals the onset of criticality~\cite{Sachdev2011}. Type-I dynamical quantum phase transitions occur far from equilibrium and are not described by a thermodynamic free energy in the conventional sense. Nevertheless, they can be characterized through the long-time behavior of suitable physical observables, which play a role analogous to equilibrium order parameters. Corresponding dynamical response functions can then be introduced to quantify the sensitivity of these observables to changes in the quench parameter. Their system-size dependence provides a useful numerical probe of the progressive sharpening of the transition.

For the type-I DQPT in the LMG model, the conventional dynamical order parameter is the long-time averaged longitudinal magnetization,
\begin{equation}
    \overline{m}_z(h_f) = \frac{1}{T}\int_0^T m_z(t;h_f)\,\dd t,
    \label{eq:average_magnetization}
\end{equation}
where
\begin{equation}
    m_z(t;h_f) = \frac{\expval{J_z(t)}}{j}
\end{equation}
is the intensive magnetization. The averaging time $T$ is chosen sufficiently long to suppress transient effects. Physically, $\overline{m}_z$ measures the average orientation of the collective spin following the quench and has been extensively employed as a dynamical order parameter in the LMG model~\cite{Goes2020,Bento2024,Nascimento2024,Nascimento2025}.

\begin{figure}
    \centering
    \includegraphics[width=1\linewidth]{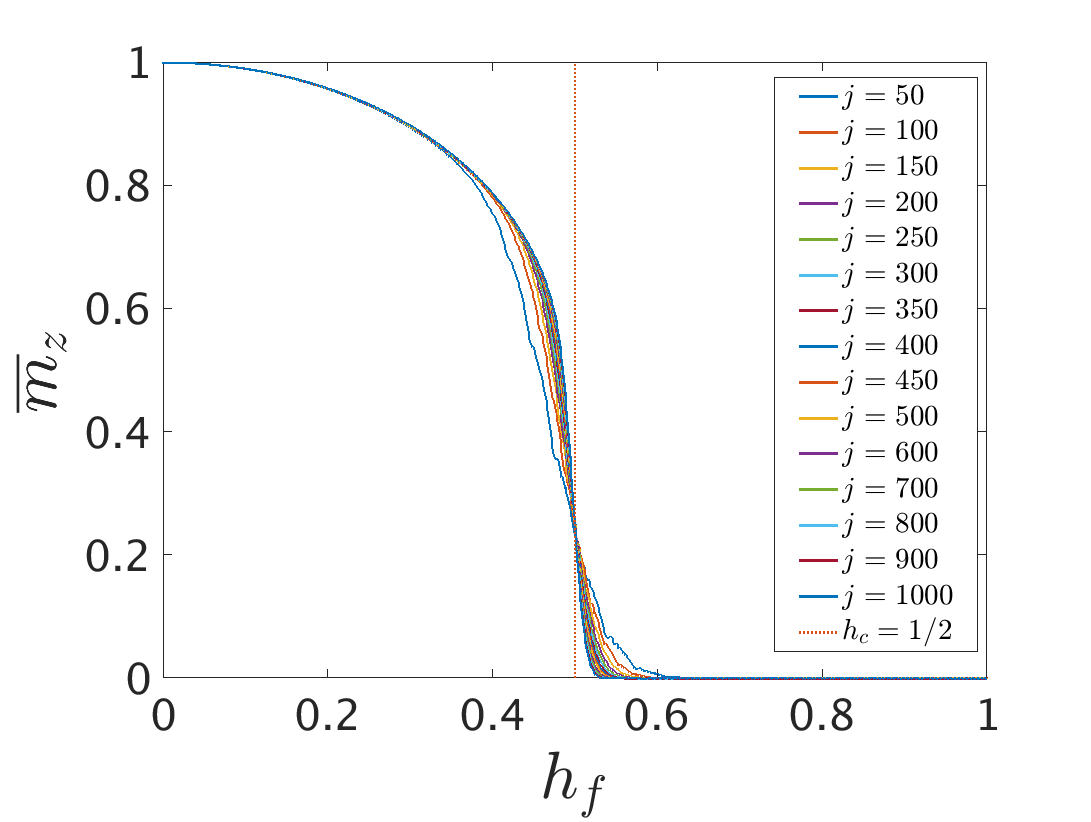}\\[2mm]
    \includegraphics[width=1\linewidth]{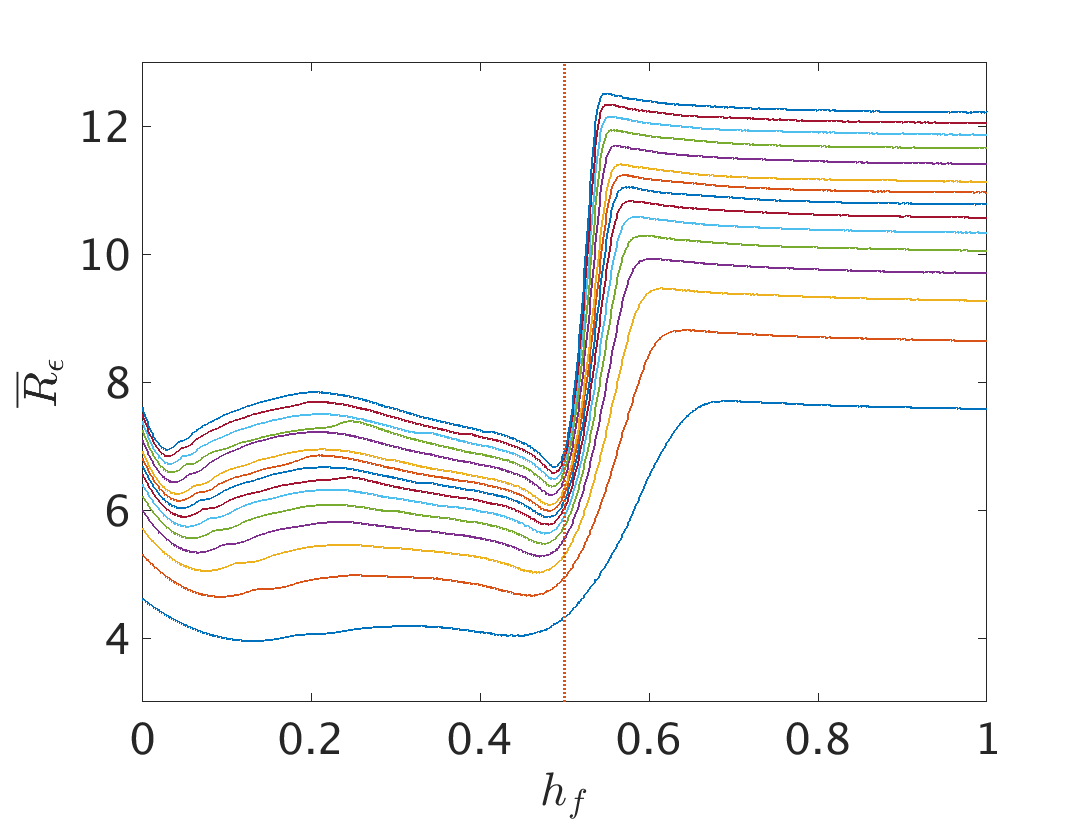}
    \caption{Long-time averaged magnetization (upper panel) and long-time averaged pre-quench energy basis rugosity (lower panel) for the LMG model as functions of the post-quench field $h_f$. We set $h_0=0$ and $\delta=1$, for which the type-I dynamical critical field is $h_c^{\mathrm{DQPT}}=1/2$, indicated by the vertical dotted red line in both panels. The same values of the total angular momentum $j$ are used in both panels. The time averages were evaluated using $T\approx 10^{4}\nu^{-1}$, where $\nu$ is the smallest Bohr frequency in the range of $h_f$ and $j$ considered.}
    \label{fig:order}
\end{figure}

The upper panel of Fig.~\ref{fig:order} shows $\overline{m}_z$ as a function of the post-quench field $h_f$ for several values of the total angular momentum $j$. For $h_f<h_c^{\mathrm{DQPT}}$ the long-time averaged magnetization remains finite. In this regime, the dynamics preserve a memory of the initial symmetry-broken state. For $h_f>h_c^{\mathrm{DQPT}}$ the average magnetization tends to zero, indicating that the dynamics explore both symmetry-related sectors and dynamically restore the broken symmetry. As $j$ increases, the crossover between these two regimes becomes progressively sharper.

Since a type-I DQPT is accompanied by a profound restructuring of the evolving quantum state, it is natural to ask whether a quantity that directly quantifies quantum-state texture may also exhibit signatures of the transition. We then consider the long-time averaged rugosity evaluated in the pre-quench energy eigenbasis,
\begin{equation}
    \overline{R}_{\epsilon}(h_f) = \frac{1}{T} \int_0^T R_{\epsilon}(t;h_f)\,\dd t.
    \label{eq:average_rugosity}
\end{equation}
This quantity characterizes the average structural complexity developed by the evolving quantum state when expressed in the eigenbasis of the pre-quench Hamiltonian. Unlike the magnetization, which probes the average polarization of the collective spin, the rugosity quantifies how strongly the state departs from the flat reference state in the chosen basis. It therefore provides complementary information about the nonequilibrium redistribution of the quantum state over the pre-quench basis.

The lower panel of Fig.~\ref{fig:order} shows $\overline{R}_{\epsilon}$ as a function of $h_f$. Below the critical field, the average rugosity remains comparatively small. When the quench crosses the critical region, $\overline{R}_{\epsilon}$ changes rapidly and approaches a higher approximately stationary value. Although the instantaneous rugosity remains oscillatory, the system spends a larger fraction of its evolution in states that are far from the flat reference state.

This behavior is consistent with the semiclassical picture discussed previously. Above the critical quench, the energy injected into the system exceeds the ESQPT separatrix, allowing the dynamics to explore both symmetry-broken sectors. The resulting redistribution of the state, together with the coherent evolution of the corresponding amplitudes, is reflected in the pronounced change of rugosity in the pre-quench energy basis. The numerical results therefore suggest that the pre-quench energy basis rugosity exhibits an order-parameter-like signature of the type-I dynamical quantum phase transition.

To characterize more quantitatively the sharpening of the transition, we introduce response functions associated with these two time-averaged quantities. In analogy with the magnetic susceptibility of equilibrium systems, we define the magnetization response as
\begin{equation}
    \chi_M(h_f) = - \pdv{\overline{m}_z}{h_f}.
    \label{eq:susc_mag}
\end{equation}
The minus sign is included so that the response is positive in the critical region, since $\overline{m}_z$ decreases as the post-quench field increases. Physically, $\chi_M$ quantifies the sensitivity of the long-time magnetization to infinitesimal changes in $h_f$.

\begin{figure}
    \centering
    \includegraphics[width=1\linewidth]{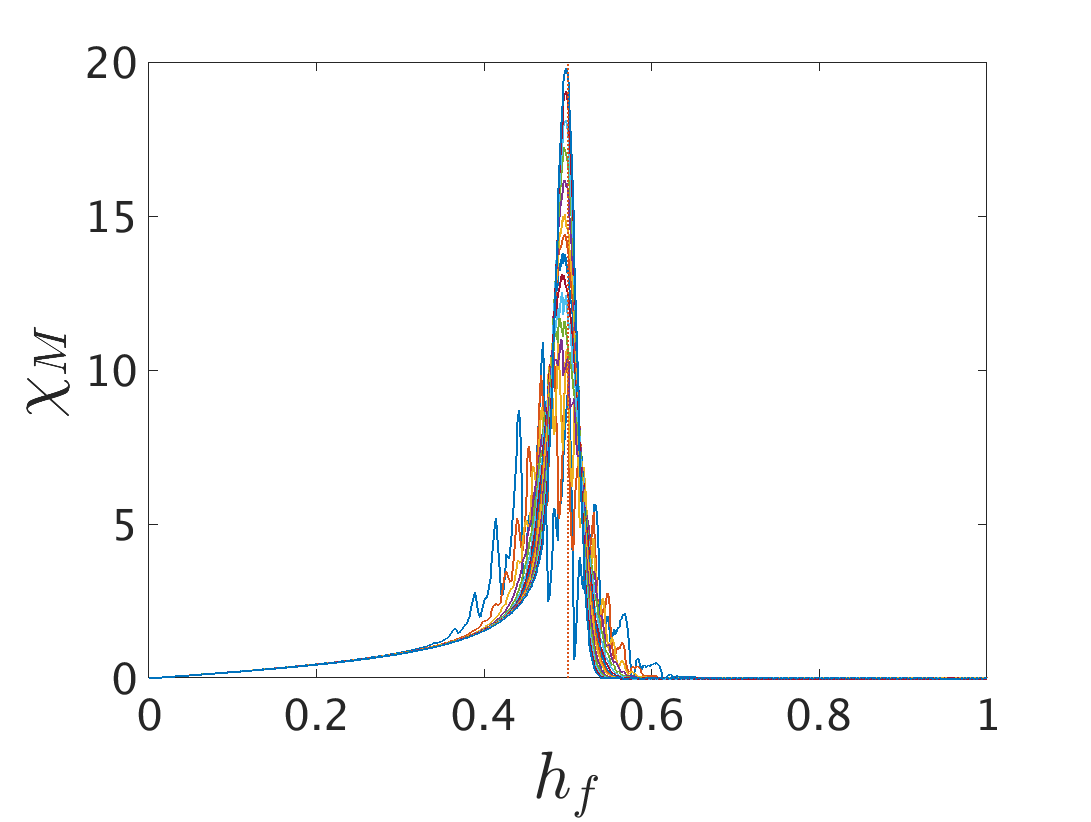}\\[2mm]
    \includegraphics[width=1\linewidth]{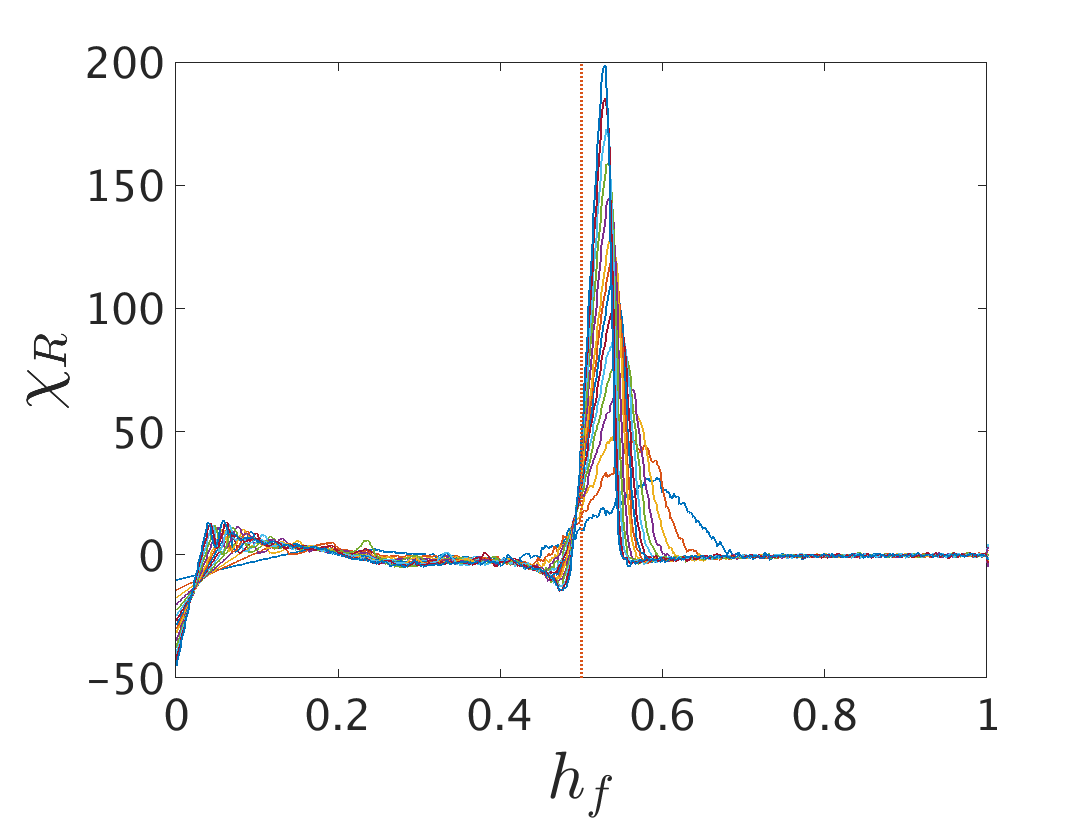}
    \caption{Response functions for the LMG model. The upper panel shows the magnetization response $\chi_M$, defined in Eq.~\eqref{eq:susc_mag}, while the lower panel shows the rugosity response $\chi_R$, defined in Eq.~\eqref{eq:susc_rug}. The derivatives were obtained numerically from the time-averaged data displayed in Fig.~\ref{fig:order}. The vertical dotted line marks the thermodynamic dynamical critical field $h_c^{\mathrm{DQPT}}=1/2$.}
    \label{fig:responses}
\end{figure}

For finite $j$, the sharp transition of the thermodynamic limit is rounded into a smooth crossover. Accordingly, $\chi_M$ remains finite and develops a maximum in the critical region. The upper panel of Fig.~\ref{fig:responses} shows that this maximum becomes progressively higher and narrower as $j$ increases, reflecting the increasing sharpness of the magnetization crossover.

Motivated by the behavior of the time-averaged rugosity across the transition, we introduce the corresponding response function
\begin{equation}
    \chi_R(h_f) = \pdv{\overline{R}_{\epsilon}}{h_f}.
    \label{eq:susc_rug}
\end{equation}
This quantity measures the sensitivity of the average rugosity to changes in the post-quench field. We use it to test whether the restructuring of the state in the pre-quench energy basis develops the same qualitative finite-size signatures as the conventional magnetization response.

The lower panel of Fig.~\ref{fig:responses} shows that $\chi_R$ develops a pronounced maximum in the critical region. As in the magnetization case, the maximum grows and becomes narrower as the system size increases. The rugosity response therefore exhibits the same qualitative finite-size sharpening associated with the dynamical transition.

To quantify the growth of the response functions, for each value of $j$ we define
\begin{equation}
    \chi_X^{\max}(j) = \max_{h_f}\chi_X(h_f,j), \qquad X=M,R.
    \label{eq:response_maximum}
\end{equation}
To quantify the increase of the response maxima with system size, we test whether the numerical data are compatible with an algebraic dependence of the form
\begin{equation}
    \chi_X^{\max}(j) \sim j^{\kappa_X}, \qquad X=M,R,
    \label{eq:peak_scaling}
\end{equation}
where the exponents $\kappa_M$ and $\kappa_R$ are obtained from least-squares fits of the numerical data.

A positive fitted exponent indicates that the response grows systematically with system size. Together with the narrowing of the response peak, this provides quantitative evidence that the smooth finite-size crossover becomes progressively sharper as the thermodynamic limit is approached. We emphasize that Eq.~\eqref{eq:peak_scaling} is used here as an empirical characterization of the numerical data, rather than being assumed a priori from an equilibrium scaling hypothesis.

\begin{figure}
    \centering
    \includegraphics[width=1\linewidth]{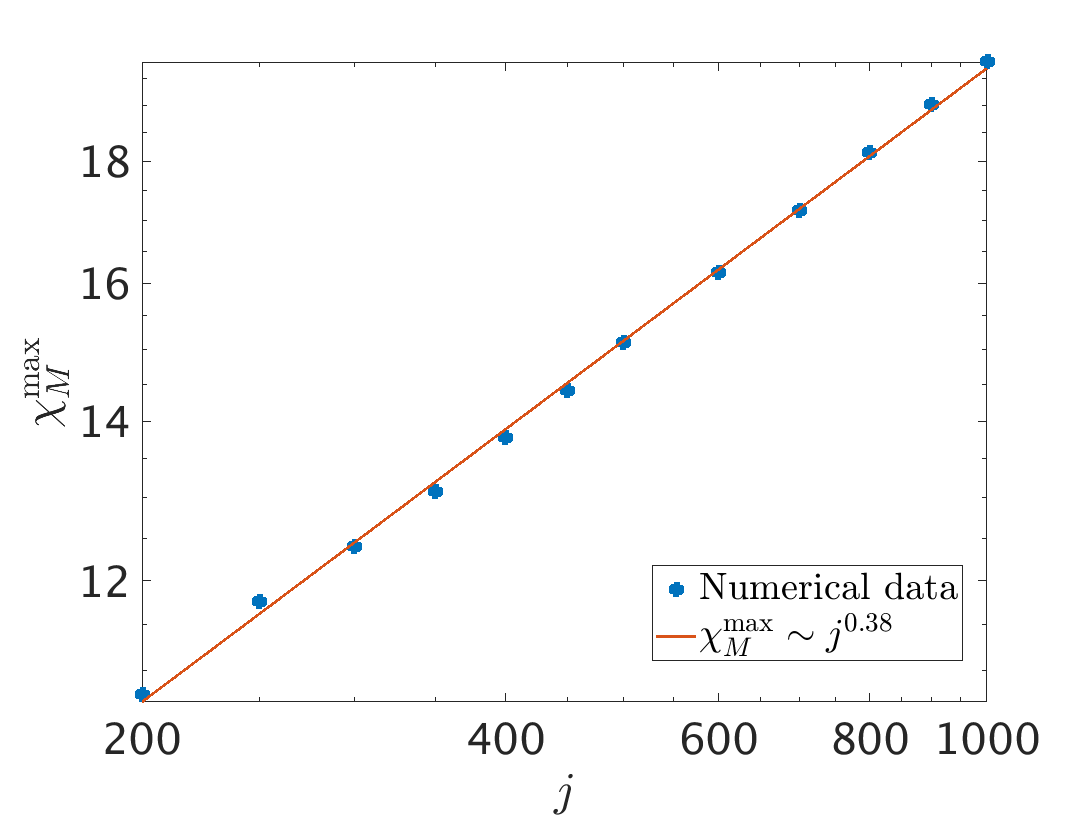}\\[2mm]
    \includegraphics[width=1\linewidth]{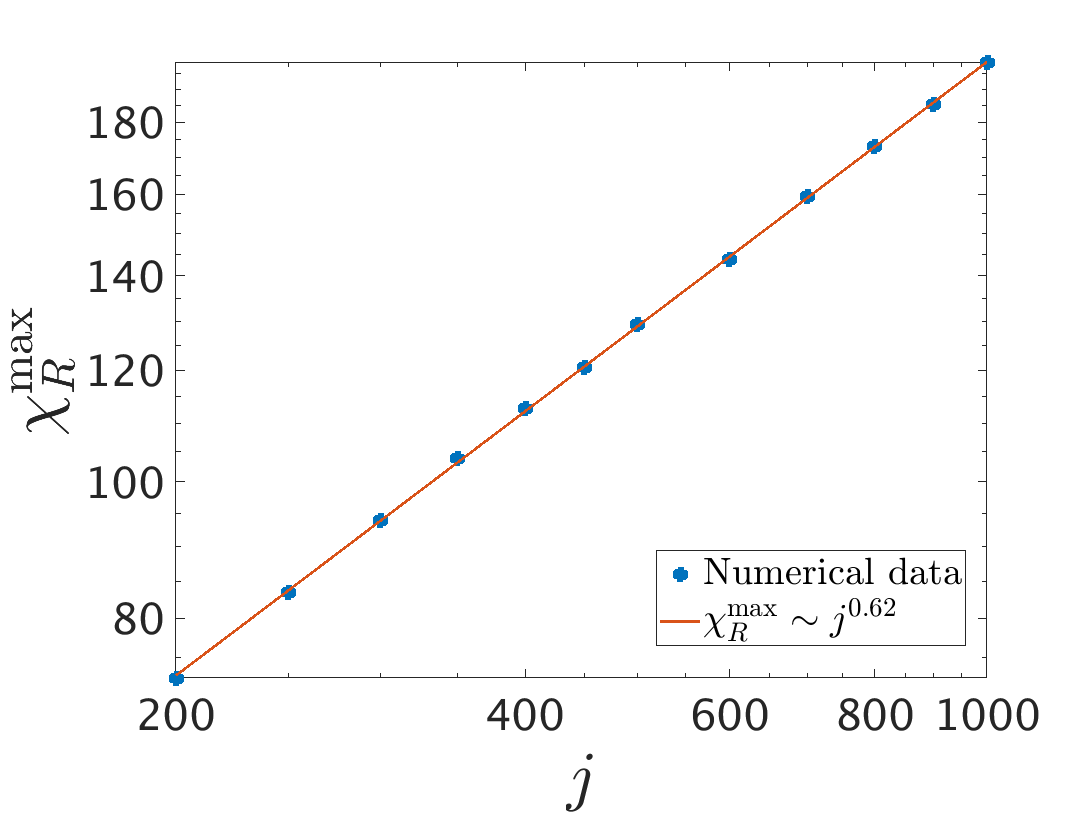}
    \caption{System-size dependence of the maximum response functions for the LMG model. The upper panel shows $\chi_M^{\max}$, while the lower panel shows $\chi_R^{\max}$. The symbols denote the numerical results for $200\leq j\leq1000$, and the lines represent fits to the algebraic form in Eq.~\eqref{eq:peak_scaling}. The fitted exponents are $\kappa_M\simeq0.38$ and $\kappa_R\simeq0.62$. The plots are shown in logarithm scale.}
    \label{fig:finite_size}
\end{figure}

Figure~\ref{fig:finite_size} shows the system-size dependence of the response maxima for some values of the angular momentum. Over the considered range, both quantities are accurately described by algebraic laws. The magnetization response yields $\kappa_M\simeq0.38$, whereas the rugosity response yields $\kappa_R\simeq0.62$.

The algebraic increase of $\chi_R^{\max}$ provides quantitative finite-size evidence that the rapid change in the time-averaged rugosity is associated with the dynamical critical behavior, rather than being merely a smooth variation with the quench field. The rugosity response exhibits a larger growth exponent over the investigated range of system sizes, indicating that it develops a particularly strong finite-size signature of the transition.

Because the magnetization and rugosity have different definitions and normalizations, the two exponents should not be interpreted as directly corresponding to the same equilibrium susceptibility exponent. Their comparison instead quantifies how the two different response functions sharpen with increasing system size.

Taken together, the behavior of the long-time averaged rugosity, the development of a pronounced response peak, and the algebraic growth of its maximum provide mutually consistent evidence that the pre-quench energy basis rugosity captures the critical restructuring of the quantum state across the type-I DQPT in the LMG model.

It is worth emphasizing that the dynamical critical point does not induce rugosity. Rather, it controls the rate at which rugosity is generated and accumulated during evolution. Near the transition, the dynamics become especially efficient at redistributing the state over the pre-quench basis, thereby enhancing the sensitivity of rugosity to the underlying critical behavior. After the critical field is crossed, the dynamics rapidly drives the system toward highly textured states, thereby accelerating the generation of rugosity. In this regime, the state spends most of its evolution far from the flat reference state, so the generation of rugosity becomes faster, causing the system to spend a larger fraction of its evolution in highly textured states.

Let us now turn to the case of type-II dynamical quantum phase transitions. At first sight, rugosity and the Loschmidt rate function appear to be unrelated quantities. The Loschmidt rate function is defined from the return probability of the evolving state to the initial state and is therefore basis independent. Rugosity, on the other hand, is intrinsically basis dependent, since it measures the distance between the evolving state and the flat state associated with a chosen basis. In general, the ground state of the pre-quench Hamiltonian is highly localized in its own energy eigenbasis rather than being a flat state, so there is no obvious connection between the two quantities.

The key observation is that this apparent mismatch can always be removed for pure initial states. Let
\begin{equation}
\ket{\psi_0} = \sum_{m=1}^{d}c_m \ket{\epsilon_m}
\end{equation}
be an arbitrary normalized initial state. Since all normalized vectors in a finite-dimensional Hilbert space are related by unitary transformations, there always exists a unitary operator $U$ satisfying $U\ket{\omega} = \ket{\psi_0}$, where $\ket{\omega}$ denotes the flat state associated with the basis $\{\ket{\epsilon_m}\}$.

Instead of viewing this transformation as acting on the state, it is more convenient to regard it as defining a new orthonormal basis,
\begin{equation}
\ket{\phi_k} = U\ket{\epsilon_k}.
\end{equation}
In this new basis, the initial state becomes
\begin{equation}
\ket{\psi_0} = U\ket{\omega} = \frac{1}{\sqrt d}\sum_{k=1}^{d} \ket{\phi_k},
\label{eq:flat_state_new_basis}
\end{equation}
that is, the initial state is precisely the flat state associated with the basis $\{\ket{\phi_k}\}$.

For the quench protocol considered here, the initial state is an eigenstate of the pre-quench Hamiltonian. In this case, a particularly convenient explicit realization of the above construction is provided by the discrete Fourier transform,
\begin{equation}
\ket{\phi_k} = \frac{1}{\sqrt d} \sum_{m=1}^{d}e^{-im\phi_k}\ket{\epsilon_m}, \qquad \phi_k=\frac{2\pi k}{d},
\label{eq:fourier_basis}
\end{equation}
which is the choice adopted throughout this work. In the case of the LMG model with $h_0=0$, this basis coincides with the eigenbasis of the operator conjugate to $J_z$.

Since the initial state is flat in this new basis, the definition of rugosity immediately gives
\begin{equation}
R_\phi(\rho_t) = - \ln\left|\ip{\psi_0}{\psi_t}
\right|^2,
\end{equation}
and therefore
\begin{equation}
\lambda_t = \frac{1}{N} R_\phi(\rho_t),
\label{eq:rugosity_echo}
\end{equation}
that is, the Loschmidt rate function is exactly the density of rugosity.

Equation~\eqref{eq:rugosity_echo} should not be interpreted as providing a new computational method for evaluating the Loschmidt rate function. Once one recognizes that every pure state can be regarded as the flat state in a suitably constructed basis, the equivalence follows directly from the definition of rugosity. Its significance is instead conceptual. The Loschmidt rate function has traditionally been interpreted as the dynamical quantity measuring the probability for the system to return to its initial state after the quench. Equation \eqref{eq:rugosity_echo} shows that this same quantity also admits an exact interpretation within the recently introduced resource theory of quantum-state texture: it is precisely the density of the quantum resource associated with rugosity. In this sense, the nonanalyticities characteristic of a type-II dynamical quantum phase transition can also be interpreted as singular generation of quantum-state texture.

The basis leading to Eq.~\eqref{eq:rugosity_echo} is not intended to be physically distinguished. Rather, it establishes that such an exact representation always exists. A more physically relevant question is whether rugosity continues to reveal dynamical criticality when evaluated in a basis selected independently of the dynamics. We now address this question by considering the eigenbasis of the pre-quench Hamiltonian.

\begin{figure}[h!]
    \centering
    \includegraphics[width=1\linewidth]{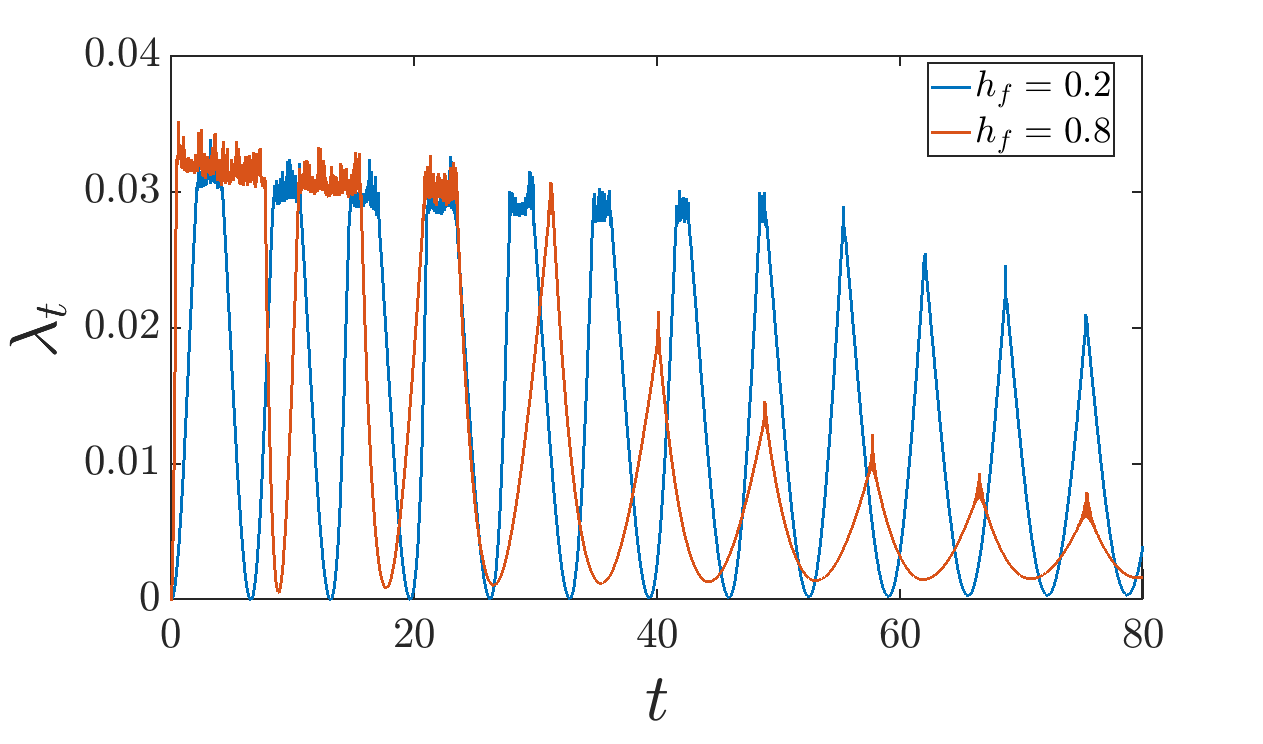} \\
    \includegraphics[width=1\linewidth]{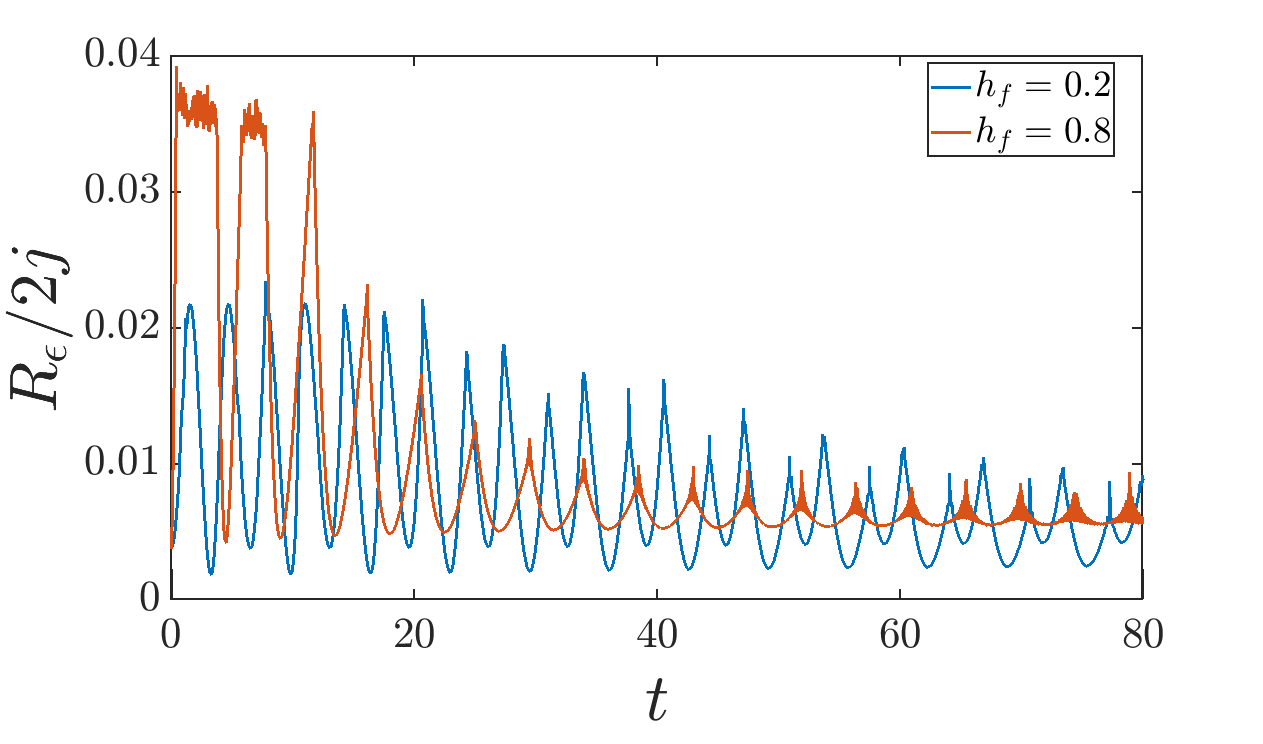}
    \caption{Rate function and rugosity density for the LMG model. The upper panel shows the rate function $\lambda_t$, which characterizes the DQPT-II, while the bottom panel shows the rugosity density $R_{\epsilon}/2j$ with respect to the pre-quench Hamiltonian eigenbasis for the same values of the quench. The fluctuations observed in the higher peaks of both panels, especially for the initial times, are due to numerical errors when computing the logarithm of very small numbers. For all plots we set $j=1000$.}
    \label{fig:rate}
\end{figure}

Figure~\ref{fig:rate} compares the rugosity computed in this pre-quench energy basis with the Loschmidt rate function for two different quenches, one crossing the dynamical critical point and the other remaining within the same dynamical phase. In this basis, rugosity is no longer identical to the rate function, since the initial state is not flat. Nevertheless, the two quantities display the same qualitative distinction between critical and noncritical quenches. In particular, the rugosity exhibits a pronounced change in its temporal behavior when the quench crosses the dynamical critical point, providing a clear signature of the transition despite the absence of the exact identity of Eq.~\eqref{eq:rugosity_echo}. This demonstrates that rugosity can reveal dynamical criticality even in physically motivated bases that are not constructed to maximize the correspondence with the Loschmidt echo.

\section{\label{sec:discussion}Discussion}

The results presented here place rugosity alongside several quantities that have been used to characterize dynamical quantum phase transitions in the LMG model. Previous studies have shown that the dynamical transition is accompanied by changes in Krylov complexity, inverse participation ratios and Shannon entropy~\cite{Bento2024}, Wehrl and thermodynamic entropies~\cite{Goes2020,Nascimento2024}, and asymmetry measures~\cite{Nascimento2025}. These quantities characterize different aspects of the reorganization of the quantum state following a quench. It is therefore important to clarify what information is specifically contained in rugosity and how it differs from these previously studied diagnostics.

A particularly useful comparison can be made with probability-based quantities such as the Shannon entropy and participation ratios. For a pure state expressed in a given basis, these quantities depend only on the occupation probabilities, whereas rugosity also depends on the relative phases of the corresponding amplitudes. A simple example is provided by the two eigenstates of $\sigma_x$. In the $\sigma_z$ eigenbasis they have identical occupation probabilities and hence the same Shannon entropy and participation ratio, while one coincides with the corresponding flat state and has zero rugosity, whereas the other is orthogonal to it and has divergent rugosity. Thus, rugosity is not a monotonic function of the Shannon entropy and cannot, in general, be reconstructed from probability-based measures. It probes instead the coherent organization of the amplitudes relative to the flat reference state.

This distinction also clarifies the relation with the other diagnostics considered in the LMG model. Krylov complexity and participation ratios characterize the spreading of the evolving state over a given representation~\cite{Bento2024}, while Wehrl and thermodynamic entropies characterize spreading, mixing, and information loss~\cite{Goes2020,Nascimento2024}. Asymmetry, in turn, quantifies coherence relative to a specified symmetry transformation~\cite{Nascimento2025}. Rugosity characterizes a different property: the coherent departure of the state from a basis-dependent flat reference configuration. These quantities should therefore be regarded as complementary diagnostics rather than alternative measures of the same physical property.

This difference is particularly relevant to the pre-quench energy basis considered here. In the LMG model, crossing the ESQPT separatrix allows the dynamics to connect the two symmetry-related semiclassical sectors, producing both a redistribution of populations and a reorganization of relative phases in this basis. The behavior of the time-averaged rugosity should therefore not be interpreted merely as a consequence of increased delocalization. Rather, it reflects the coherent reorganization of the state relative to the flat reference configuration selected by the pre-quench Hamiltonian.

Finally, the basis dependence of rugosity is essential to distinguish the two results obtained in this work. For type-II DQPTs, the basis is constructed so that the initial state is flat, yielding an exact change-of-representation identity between the rugosity density and the Loschmidt rate function. For the type-I DQPT studied in the LMG model, by contrast, the pre-quench energy eigenbasis is fixed by the initial Hamiltonian and is not selected to optimize the dynamical signature. In this physically motivated basis, rugosity exhibits an order-parameter-like behavior and a pronounced finite-size critical response. Thus, rugosity does not replace existing diagnostics, but provides complementary, phase-sensitive information about the reorganization of the quantum state across the dynamical transition. Whether analogous signatures arise in other many-body systems and in other physically motivated bases remains an open question.

\section{\label{sec:conclusion}Conclusions}

In this work, we investigated the connection between dynamical quantum phase transitions and rugosity as a measure of quantum-state texture. For type-I DQPTs, we focused on the Lipkin-Meshkov-Glick model and showed that the time-averaged rugosity evaluated in the eigenbasis of the pre-quench Hamiltonian exhibits an order-parameter-like behavior across the dynamical transition. This interpretation is further supported by the associated response function, whose peak becomes progressively sharper with increasing system size and whose maximum displays algebraic growth over the range of sizes investigated. Within the LMG model, this behavior can be understood from its semiclassical structure: crossing the ESQPT separatrix changes the dynamical connection between the two symmetry-related sectors and reorganizes the state in the pre-quench energy basis. For type-II DQPTs, we showed that, for any pure initial state, one can construct a basis in which that state is flat. In this representation, the rugosity density coincides exactly with the Loschmidt rate function. This result should be understood as an exact change-of-representation identity rather than as the selection of a uniquely preferred physical basis.

These two results highlight distinct roles of rugosity in dynamical critical phenomena. In the type-II construction, the basis is chosen so that the rugosity reproduces the Loschmidt rate function exactly. In the type-I analysis of the LMG model, by contrast, the pre-quench energy basis is fixed independently by the initial Hamiltonian, and rugosity provides a physically motivated signature of the transition without being trivially equivalent to a conventional DQPT diagnostic. As discussed above, rugosity also contains information that is not generally encoded in probability-based quantities such as Shannon entropy or participation ratios, since it is sensitive to the relative phases of the basis amplitudes. It should therefore be regarded as complementary to existing diagnostics such as complexity, entropy, and asymmetry, rather than as a replacement for them.

Several questions remain open. Most importantly, the type-I results established here are restricted to the LMG model. Whether analogous rugosity signatures occur in systems without a semiclassical double-well structure or an ESQPT separatrix, such as short-range spin chains and fermionic lattice models, remains to be investigated. More generally, the basis dependence of rugosity raises the question of which physically motivated representations are most informative for a given nonequilibrium process and how robust the corresponding signatures are under changes of basis. Extensions to mixed states, open quantum systems, and finite-temperature dynamics provide further natural directions. Exploring these questions may clarify the extent to which quantum-state texture provides a useful information-theoretic perspective on nonequilibrium quantum criticality.

\begin{acknowledgments}
LCC acknowledges support from Brazilian National Council for Scientific and Technological Development (CNPq) through grant 308065/2022-0, the National Institute of Science and Technology for Applied Quantum Computing through CNPq grant 408884/2024-0, Goiás Research Foundation (FAPEG) through grant 202510267001843, and São Paulo Research Foundation (FAPESP) through grant 2025/23726-4. LCC also acknowledges the warm hospitality of the Instituto de Física de São Carlos. K.Z. acknowledges CNPq (Grant No.305665/2025-1). I.M. acknowledges financial support from  FAPESP (Grant No. 2022/08786-2). D.O.S.P. acknowledges support from CNPq (Grant No. 305766/2026-0).  
\end{acknowledgments}


\end{document}